\documentclass[onecollarge,runningheads]{svjour2}
\smartqed  % flush right qed marks, e.g. at end of proof
\usepackage{graphicx}
\journalname{Journal of Biological Physics}
\begin{document}

\title{Dynamics of lysozyme and its hydration water under electric field%\thanks{Grants or other notes
}


\author{P.M. Favi,      \and
        Q. Zhang, 		\and
        H. O'Neill,		\and
        E. Mamontov,	\and        S.O. Diallo}


\institute{P.M. Favi \at
           Quantum Condensed Matter Division, Oak Ridge National Laboratory, Oak Ridge, Tennessee 37831, USA  
            \emph{Permanent address:} Department of Materials Science and Engineering, University of Tennessee, Knoxville, Tennessee 37996, USA            \and
           Q. Zhang and H. O'Neill \at
              Biology and Soft Matter Division, Oak Ridge National Laboratory, Oak Ridge, Tennessee 37831, USA    
            \and
            E. Mamontov \at
             Chemical and Engineering Materials Division, Oak Ridge National Laboratory, Oak Ridge, Tennessee 37831, USA  
            \and
            S.O. Diallo  \at
            Quantum Condensed Matter Division, Oak Ridge National Laboratory, Oak Ridge, Tennessee 37831, USA  
                       \email{omardiallos@ornl.gov}           \\ 
}

\date{Received: \today / Accepted: date}

\maketitle

\begin{abstract}
The effects of static electric field on the dynamics of lysozyme and  its hydration water have been investigated by means of incoherent quasi-elastic neutron scattering (QENS).  Measurements were performed on lysozyme samples, hydrated respectively with heavy water (D$_2$O) to capture the protein dynamics, and with light water (H$_2$O), to probe the dynamics of the hydration shell, in the temperature range from 210 $<T<$ 260 K.  The hydration fraction in both cases was about $\sim0.38$ gram of water per gram of dry protein. The field strengths investigated were respectively 0 kV/mm and 2 kV/mm  ($\sim2\times10^{6}$ V/{m}) for the protein hydrated with D$_2$O and 0 kV and 1 kV/mm for the H$_2$O-hydrated counterpart.  While the overall internal protons dynamics of the protein appears to be unaffected by the application of electric field up to 2 kV/mm, likely due to the stronger intra-molecular interactions,  there is also no appreciable quantitative enhancement of the diffusive dynamics of the hydration water, as would be anticipated based on our recent observations in water confined in silica pores under field values of $~2.5$ kV/mm. This may be due to the difference in surface interactions between water and the two adsorption hosts (silica and protein), or to the existence of a critical threshold field value $E_c\sim2-3$ kV/mm for increased molecular diffusion, for which electrical breakdown is a limitation for our sample.

\end{abstract}

\section{Introduction}
\label{intro}
Interactions of proteins with charged surfaces are important in many applications such as chromatographic separation \cite{Roth:98}, biosensors \cite{Talasaz:06}, and design of biocompatible medical implants \cite{Subrahmanyam:02}. Knowledge of the interactions involving charged, polar and polarizable groups, and the hydration water in proteins is thus of fundamental interests because it provides microscopic insights into biophysical molecular recognition and protein folding mechanism \cite{Ojeda:10}.  

Many opened questions remain regarding the exact way in which proteins respond to external stress, including electric field. Since the function of proteins is critically linked to their three-dimensional structures and to their hydration water, exposure to any form of stress, thermal or non, which may induce changes in conformation can alter cellular function. This is particularly relevant in today's environment, with the increasingly use of portable electromagnetic devices, which raises many questions about the possible effects on human health. 

There is strong evidence that protein misfolding is responsible for a number of known diseases such as prion diseases, Alzheimer's disease, and so on \cite{Ojeda:10,Lange:09}. Investigating the effects of the different contributing factors (such as temperature, hydration, pressure, pH, ionic strengthÉ) on the conformational changes of proteins is thus key to understanding this complex mechanism. Application of an electric field is such a parameter since the protein itself has an internal dipole moment. The force exerted on the protein dipole will result in a torque that will rotate the protein and is likely to affect the diffusion motions of the hydration water, which are known to be on the nano- to pico-second time scales \cite{Teixeira:85}. Charged and polar groups are also expected to move little with the application of electric field \cite{Xu:96}.

Considerable progress has already been made in structural studies involving intermediate states along the folding pathways of some proteins under electric field and the effects on their crystallization \cite{Taleb:01,Nanev:01,Pompa:05}. This body of work is supported by several molecular simulation techniques \cite{Marklund:09,Abrikossov:10,Sun:89,Budi:07}. Similarly, but to a lesser extent, there has been a variety of spectroscopy techniques, principally H-D exchange NMR, circular dichroism, and fluorescence spectroscopy and protein engineering, that have provided some information about the changing environments and configurations of individual residues during the folding process \cite{Lange:09,Pompa:05}. Unfortunately, these measurements are not able to resolve the spatial differences in the dynamics of the hydration water nor are they able to dissociate the different type of motions involved. For instance; Are the molecular motions localized or long-ranged? Are they translational or rotational in nature? Such questions can uniquely be answered by quasi-elastic neutron scattering (QENS), thanks to the dependence of the observed relaxation times on the wavevector $Q$, which can not be accessed with other spectroscopy techniques such as NMR \cite{Varga:10,Roh:06}. 

\begin{figure*}
\centering
  \includegraphics[width=0.85\textwidth]{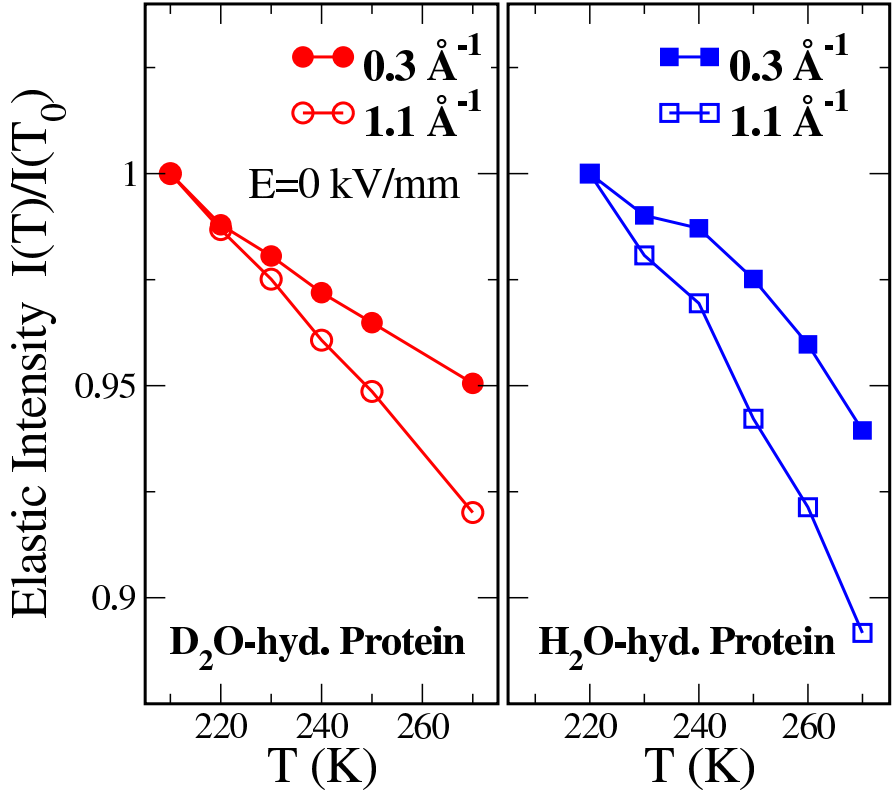}
\caption{Temperature dependence of the elastically scattered neutron intensity obtained from  lysozyme hydrated with D$_2$O (left panel) and with H$_2$O (right panel),  at selected $Q$ values of 0.3 {\AA$^{-1}$} and 1.1 {\AA$^{-1}$} in the absence of field (E$=0$ kV/mm). The stronger suppression in intensity in the H$_2$O hydrated protein arises from the larger mean-squared proton displacement $\langle r^2(T)\rangle$ in the Debye-Waller factor compared to the  D$_2$O hydrated sample. The associated uncertainties on the data points are around 6-8\%, and are much smaller than the symbols.}
\label{fig:1}      
\end{figure*}

We have applied the QENS technique to study of the effects of applied electric field on the diffusive motion of hydration water in a model protein system, Hen-Egg-White-Lysozyme (HEWL), a residue protein found in secretions (e.g., saliva, sweat, and mucus), well-known for  its protective property against certain bacterial or virus aggressions.  We find no effect of external electric field on the dynamics of Lysozyme, probably due to to its stronger intra-molecular interactions \cite{Bon:02}. Similarly, but in contrast to our recent observations in water confined in silica pores under field values of $2.5$ kV/mm \cite{Diallo:12}, we observe no appreciable increase in the diffusive dynamics of the hydration water under field. \cite{Diallo:12,Vegiri:02}

\section{Experimental Details}\label{sec:1}
\subsection{Sample Preparation}
The lysozyme sample was purchased from Sigma Aldrich (L4919; 98\%purity). This commercial batch was used without further purification.We prepared two samples for the neutron scattering measurements,  one hydrated with H$_2$O and another with D$_2$O \cite{Mamontov:10}.  The samples were lyophilized repeatedly before being hydrated. In both cases, the labile hydrogen atoms were exchanged for deuterium atoms by dissolving in heavy water (D$_2$O), prior to lyophilization. The samples were thus hydrated using isopiestic conditions by incubation in a sealed container containing respectively 99.9\% of H$_2$O and 99.9\% of D$_2$O. The level of hydration was controlled by varying the incubation time. The final hydration level $h$ was determined by the relative change in the sample weight following humidity exposure, yielding an $h\simeq$38\% for each sample.  Neutron-scattering measurements were performed on the backscattering silicon spectrometer (BASIS) at the 1 MW Spallation Neutron Source, Oak Ridge National Laboratory (ORNL), USA \cite{Mamontov:11}. 

Each protein sample was mounted between two Al plates separated with a Teflon gasket (with $\sim$1 mm gap), specially designed for high voltage experiments. The assembly was then sealed using Teflon based screws and a VITON O-ring, which are shielded out of the beam with Boron Nitride mask. The cell assembly is then mounted on the cold head of a standard closed-cycle refrigerator, and to a 10 kV TREK power supply  for the voltage application. 

\begin{figure}
\centering
  \includegraphics[width=0.85\textwidth]{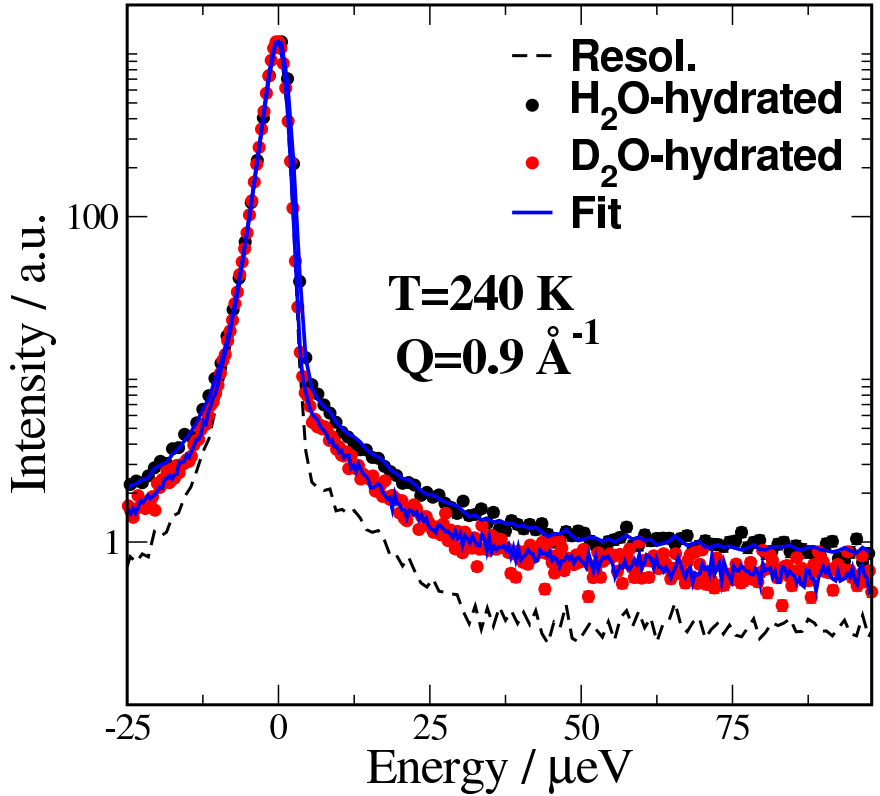}
\caption{Superposition of the QENS response obtained from the protein hydrated with H$_2$O (black solid circles) and D$_2$O (red solid circles), at temperature $T=$240 K and $Q$=0.9 {\AA$^{-1}$}. The larger broadening in the H$_2$O-protein sample is due to the hydration shell. The corresponding fits are shown as solid blue lines and the instrument resolution (measured at 50 K using the exact same sample) is shown in dashed line for comparison.}
\label{fig:2}       % Give a unique label
\end{figure}

\subsection{Neutron scattering}
QENS is a powerful spectroscopy technique for determining dynamics in hard and soft matter.  Alternative and complementary techniques to QENS include Raman spectroscopy, far infrared spectroscopy, NMR and so on. However, the unique advantage of neutron scattering in biological systems arises from its different sensitivity to light atoms such as deuterium and hydrogen, which results in remarkable differences between samples containing different proportion of these isotopes. Moreover, unlike other standard techniques, QENS uniquely probes geometrical information associated with molecular dynamics. Generally, the observed  quantity in a neutron scattering measurement is the dynamical structure factor $S({\bf Q},E)$ (or DSF), which is related to the probability for an incident neutron to be scattered with a wavevector transfer  ${\bf Q}$ and an energy transfer  $E$ to the sample. The $E$ and ${\bf Q}$-dependence of $S({\bf Q},E)$ provides information on the characteristic correlation times $(t\sim\hbar/E)$ and on the geometry ($r\sim1/Q$) of the molecular motions within the sample, respectively.  For an isotropic system, as is the case here, the DSF allows for a \lq powder averaging' of the signal, and consequently it depends effectively on the magnitude $Q$ of the wavevector transfer rather than on the vector ${\bf Q}$. In most cases, the DSF contains both coherent and incoherent scattering contributions, arising from pair- and self-correlations, respectively. However, and thanks to the large incoherent cross-section of hydrogen over that of deuterium (and all other elements), it is possible to mask the dynamics in part of the sample with selective deuteration. In this event, the dynamics seen by the neutron comes largely from the incoherent part,  yielding information on self-diffusion processes. 

To probe such dynamics, which are typically in the pico-to-nano second regime, a state-of-the-art high energy resolution neutron scattering instrument such as BASIS \cite{Mamontov:11} is required. The wide accessible dynamic range $\Delta$E$=\pm$100 $\mu$eV combined with the excellent energy resolution of 3.5 $\mu$eV (Full-Width at Half Maximum or FWHM) at the elastic position makes BASIS an ideal spectrometer for probing the pico-nano second dynamics in lysozyme and its hydration water.  The $Q$-range investigated here varies from 0.3 to 1.1 \AA$^{-1}$ in step of  $\Delta Q=$0.2 \AA$^{-1}$.

\paragraph{Incoherent elastic signal} 
For diagnostic purposes, we perform rapid standard  \lq elastic' scans on both lysozyme samples hydrated with H$_2$O and D$_2$O. The aim was to check the quality of the neutron signal from the samples, and to provide a calibrated temperature range for the subsequent QENS measurements, which requires high counting times. Data were collected with 10 K temperature increments on cooling from 270 K to 210 K. Fig. \ref{fig:1} shows the normalized elastic intensity (with respect to the maximum lowest temperature value $I(T_0)$) as a function of temperature for the D$_2$O and H$_2$O hydrated proteins, at two selected $Q$ values, lowest and highest $Q$ values investigated. The elastic intensity for each temperature was obtained by integrating the corresponding spectrum over a very narrow energy range of ${\pm}$3.5 $\mu$eV, corresponding to the elastic resolution. For an isotropic powder sample, the elastic intensity is expected to have a Debye-Waller behavior, $I(T)\sim e^{-Q^2\langle r^2(T)\rangle/3}$, where $\langle r^2(T)\rangle$ is the mean square amplitude vibration of the molecule. As the sample cools down,  the molecular diffusion start to slow down and $\langle r^2(T)\rangle$ decreases, yielding an increase in the elastic line. The elastic intensity  within  the 3.5 $\mu$eV energy resolution  effectively increases with decreasing temperature but never reaches a maximum plateau region down to the lowest temperature investigated, suggesting that both the protein and its hydration water are still mobile below 220 K. The stronger suppression in intensity in the protein hydrated with H$_2$O with increasing temperature arises from the larger mean-squared proton displacement $\langle r^2(T)\rangle$ in the Debye-Waller factor of the hydration water in the H$_2$O hydrated protein.
 
 \begin{figure}
\centering
  \includegraphics[width=0.85\textwidth]{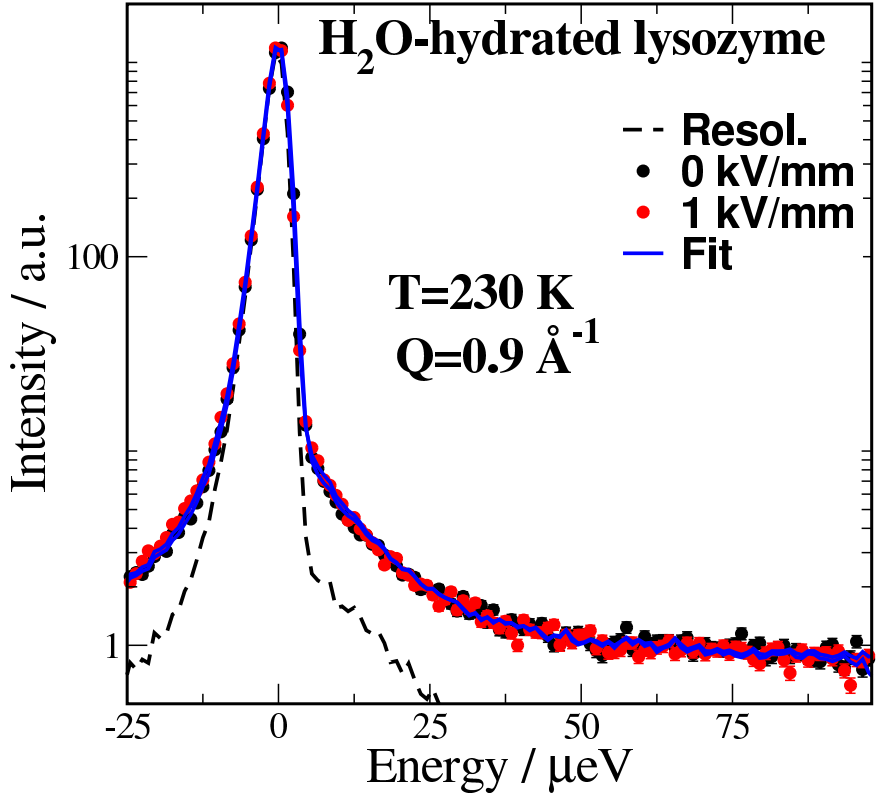}
  \caption{Field dependence of the QENS response obtained on the H$_2$O-hydrated protein at temperature $T=$230 K and wavevector $Q$=0.9 {\AA$^{-1}$}.  The black solid symbols represent the 0 kV/mm applied field data, and the red solid circles the 1 kV/mm data. The corresponding fits are shown as solid blue lines and the instrument resolution is shown as a dashed line for comparison.}
\label{fig:3}       % Give a unique label
\end{figure}

\paragraph{Quasi-Elastic Neutron Scattering} 
Generally, the observed incoherent DSF is a convolution of the translational DSF  and the rotational  one. By using only the spectra at low $Q$ values where rotations are generally not observed, (for water molecules on BASIS,  generally $Q\leq1$ {\AA}$^{-1}$) the rotational contributions can be conveniently neglected. The QENS data  were thus investigated for wavevector transfers $Q$, 0.3$\leq Q\leq$1.1 {\AA}$^{-1}$, in steps of $\Delta Q=$ 0.2 {\AA}$^{-1}$,  within the same temperatures range as the elastic scans above.  The measurements were performed with and without the application of external electric field. QENS Measurements were initially taken on the sample hydrated with D$_2$O at field values of $E=0$ and 2 kV/mm, followed by the measurements on the H$_2$O-hydrated protein sample at $E=0$ and 1 kV/mm. These field strengths are the maximum achievable values below  which electrical breakdown does not occur within our experimental set-up. The likelihood of such an event was carefully monitored with an oscilloscope via a  simultaneous measurement  of the current and voltage across the sample.   
 
\begin{figure}
%\centering
  \includegraphics[width=1.0\textwidth]{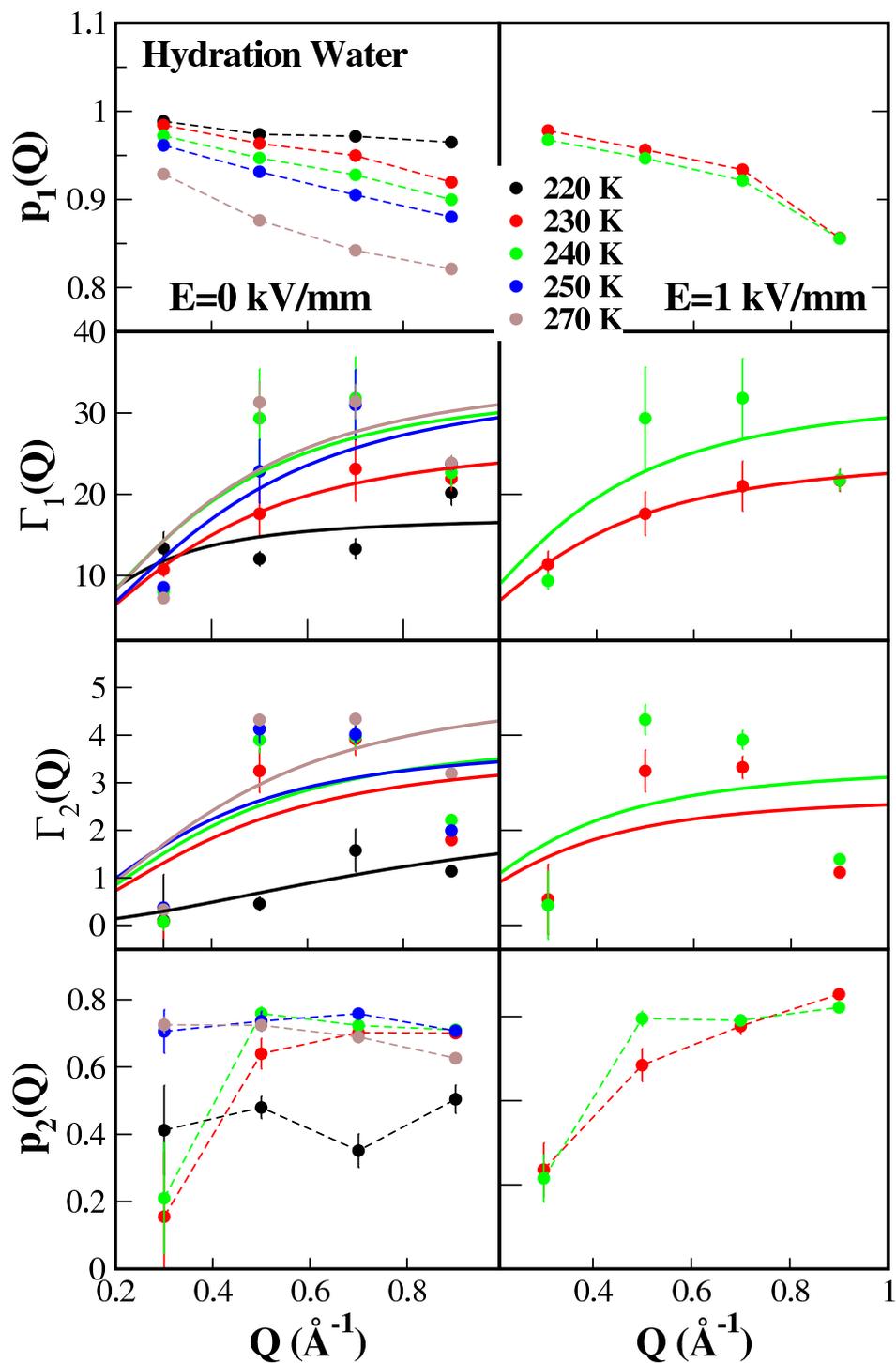}
\caption{Fit parameters characterizing the dynamics of the hydration water. Panel on the left summarizes the parameters with no field and the panel on the right the results when the field is applied. The parameter $p_1(Q)$ denotes the overall elastic fraction arising from all immobile atoms, as seen by the spectrometer window,  $\Gamma_1$ the broad Lorentzian width associated with the fast molecular dynamics, $\Gamma_2$  the narrow Lorentzian width associated with the slower molecular dynamics, and $p_2(Q)$ is the relative weight of the narrow Lorentzian, as described in the text. The relatively low quality of the fits reflects the limited statistics of our QENS data}
\label{fig:4}    
\end{figure}
\section{Analysis and Results} \label{sec:2}
To analyze the data, we begin by qualitatively comparing the different spectra collected on the sample hydrated with H$_2$O with that hydrated with D$_2$O to assess the relative contribution of the hydration water with respect to that of the protein.   Fig. \ref{fig:2} shows the observed QENS spectra at temperature $T=$ 240 K at selected $Q=0.9$ \AA$^{-1}$ in the absence of field. The overlaid solid lines are model fits, as described below.  The dashed line is the instrument resolution taken with the H$_2$O-hydrated sample at $T=50$ K, where all relevant molecular motions are frozen out.  There is clearly an appreciable excess broadening of the QENS signal in the protein hydrated with H$_2$O compared with that hydrated with D$_2$O, indicating a discernible dynamics quite different to that of the protein itself. 

\subsection{Proteins dynamics}
The internal dynamics of protein is complex, and not well captured by \lq standard' Lorentzian line fits used in QENS analysis. Rather, a stretched exponential function in the form $e^{(-\frac{t}{\tau})^{\beta}}$, with no single activation energy generally works best \cite{Roh:06}.  It is also  common to study the \lq mean' square amplitude vibration of the protons inside the proteins, which comes to a large extent from the side groups (hydrogen atoms on the methyl groups for example).  In the present case, this is reflected in the monotonic reduction of the elastic intensity with increasing temperature, as thermal fluctuations affect $\langle r^2(T)\rangle$. We would like to emphasize here that our primary objective is to study the effects of the static electric field on the hydration water. A qualitative conclusion regarding the protein dynamics can be drawn by visually inspecting the difference in the response from the D$_2$O hydrated sample when the field is applied and when it is not. Our main observation is that there is no obvious difference between the two, within the experimental precision. We conclude that the application of field (up to the maximum field applied here; i.e., $\sim$1 kV/mm) has no observable effects on the protein dynamics, on the 1 nano- to 1 pico-second time scale.  The present observations are consistent with the molecular dynamics predictions of Xu et al. \cite{Xu:96}, in which the dynamics of the protein bovine pancreatic inhibitor were investigated under a static electric field (AC). The authors found no effects of field on the protein dynamics for field values below $~10^{8}$ V/m ($\sim$ 100 kV/mm).

\subsection{Hydration water}
To uniquely characterize the dynamics  of the hydration water (and exclude the protein contributions), we subtracted the spectra of the H$_2$O-hydrated sample ($S_{L_{H}}(Q,E)$) from that of the D$_2$O hydrated protein ($S_{L_{D}}(Q,E)$) using the correct mass ratio and the relative sample transmissions, following methods described elsewhere  \cite{Orecchini:08,Orecchini:09}. The resulting spectra  represent the net signal from the hydration water, as discussed below. Based on the relative signal strength between the two samples, as highlighted for example in Fig.\ref{fig:2},  we expect the overall signal obtained on the sample hydrated with H$_2$O to be dominated by an \lq average' broad Lorentzian term. Another Lorentzian function, somewhat narrower, was required to fully reproduce the observed QENS spectra.  Within the $Q$-range investigated, the model using a double Lorentzian model captured the signal from the hydration shell data reasonably well, in agreement with previous arguments \cite{Qvist:11} and findings \cite{Diallo:12}. The dynamical structure factor for the hydration water $S_{wat}(Q,E)$ is thus approximated by,
  
 \begin{eqnarray}
S_{wat}(Q,E)&=& S_{L_{H}}(Q,E)-{\eta} S_{L_{D}}(Q,E)\\ \nonumber
                       &=& (1-p_2(Q))\frac{1}{\pi}\frac{\Gamma_1(Q)}{\Gamma_1^2(Q)+E^2}+p_2(Q)\frac{1}{\pi}\frac{\Gamma_2(Q)}{\Gamma_2(Q)^2+E^2}
\label{eq1}
\end{eqnarray}
\noindent where $p_2$ is the relative weight of the narrow component,  $\Gamma_{1,2}$ are the HWHMs associated with the dynamics of the hydration shell, $\eta$ is the ratio between the protein mass contained in the H$_2$O hydrated sample and that contained in the D$_2$O-hydrated sample scaled to the relative neutron transmission factors ($\eta\simeq$0.4).  The above fitting model is then convoluted with the instrument resolution such that the measured scattering function $I(Q,E)$ is, 
 \begin{equation}
  I(Q,E)=A(Q)\large[ p_1(Q)\delta(E)+(1-p_1(Q))S_{wat}(Q,E)\large]\bigotimes R(Q,E) +B(Q,E)
 \label{eq2}
 \end{equation}
 \noindent Here $R(Q,E)$ is the instrument resolution, measured with the H$_2$O-hydrated sample at temperature of 50 K where all observable dynamics within our sample are frozen out. The weight $p_1$ is the fraction of water molecules elastically scattered by the neutrons plus those  that appear to be \lq immobile' on our spectrometer time window.
The term $B(Q,E)=a+bE$ is a small correction required to account for the residual function left after the subtraction in Eq. \ref{eq1}. The protein-water samples are likely to lead to un-wanted multiple scattering effects, which are non-trivial to account for.  The subtraction method above, when properly done \cite{Orecchini:08}, yields a net hydration water signal with negligible contribution from multiple scattering. Any residual multiple scattering effect  would be buried under the small $Q$-independent term $a$ in the function $B(Q,E)$. 

We note that the convolution in Eq. \ref{eq2} leads to fitted lines with statistics limited by those of the measured resolution function. This effect tends to be more apparent in the background regions, as can be observed in Figs. \ref{fig:2} and \ref{fig:3}. While the data can be smoothed to improve visualization, we chose to present our data using the native binning of BASIS ($\Delta$E=0.4 $\mu$eV). 

\begin{figure}
\includegraphics[width=1\textwidth]{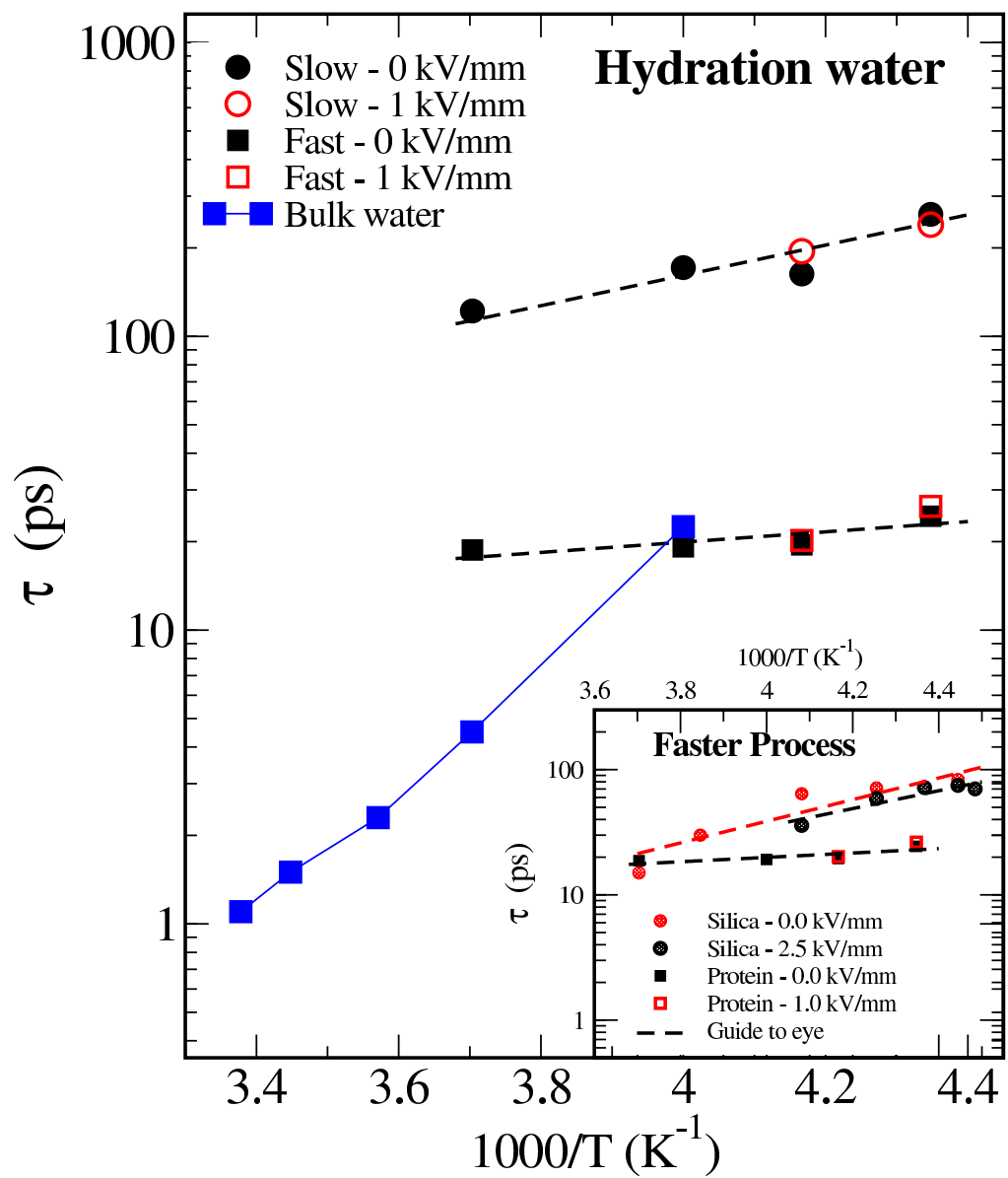}
\caption{(Color) Temperature behavior of the average residence time $\tau$ of  water surrounding lysozyme with (closed black symbols) and without field (open red symbols).  The two types of observed relaxations (fast and slow) exhibit both an Arrhenius temperature behavior (i.e. $e^{-\frac{E_a}{RT}}$). No significant effect of the field is observed. Bulk water data (open blue squares) are shown for comparison \cite{Teixeira:85}. The inset compares the $\tau$ of the fast component observed here (open and solid squares) with those of water adsorbed in tight silica pores (filled dotted circles), as reported in Ref. \cite{Diallo:12}. }
\label{fig5}
%\label{fig6}
\end{figure} 
 
 We attribute the broader of the two Lorentzians to the  \lq caged' motion of  water molecules (fast transient motion inside a molecular cage), and the narrow component to the \lq cage-breaking' water molecules which diffuse comparatively slower \cite{Diallo:12,Qvist:11}. Recent compelling arguments for using Eq. \ref{eq1} to describe the dynamics of confined water can be found in Ref. \cite{Qvist:11}. To illustrate the quality of the fits obtained with Eqs. \ref{eq1} and \ref{eq2}, we show (as an example) the fits obtained at selected temperature and $Q$ values, as solid lines in Fig. \ref{fig:2} and Fig.\ref{fig:3}.  A summary of the temperature dependence, as well as the wave-vector dependence of the fit parameters is shown in Fig. \ref{fig:4}.  The observed $\Gamma_{1,2}(Q)$ at each temperature and field value were fit using the following jump diffusion expression, 
\begin{equation}
\Gamma_{1,2}(Q)=\hbar\frac{D_{1,2}Q^2}{1+D_{1,2}{Q{^2}}\tau_{1,2}}
\label{eq3}
\end{equation}
\noindent to determine the average residence time $\tau$ between conformation jumps, and  $D$ the corresponding diffusion coefficients. The subscript denotes whether it is the fast (1) or slow dynamics (2).  Fig. \ref{fig:4}. From  the $D=\langle r^2\rangle/6\tau$, it is possible to estimate the mean squared diffusion jump length $\langle r^2\rangle$.  To a good approximation, the diffusion coefficient $D$ is behaves as $\hbar Q^2$ at low $Q$, and as $\hbar/\tau$ at the higher $Q$ values. The fit results are summarized in Table \ref{tab:1}. 

In summary, the present QENS study reveals two type of motions for the hydration water (a fast and slow motion). Both processes are very much temperature dependent,  increasing with increasing temperature with an Arrhenius behavior. To quantify this temperature evolution, we first attempted to determine the diffusion parameter $D$ at each temperature $T$. To a first approximation, $D$ can be obtained from a series expansion of Eq. \ref{eq3} around $Q\simeq0$, a limit in which $\Gamma\simeq \hbar D Q^2$, as explained above. Unfortunately, $D$ is simply not a well determined fit parameter due to significant small angle scattering signal at low $Q$. This limitation is not specific to our system, but quite commonly encountered with powder samples. The parameter $\tau$ on the other hand is a well determined quantity in the high $Q$ limit. Moreover because the two identified dynamics processes are well separated in time scales (1-2 order of magnitude), the corresponding relaxation times ($\tau_1$ and $\tau_2$) are clearly distinguishable on BASIS, as depicted in Fig. \ref{fig5}.   Overall however, the water relaxation processes are largely unaffected by the application of external field (up to the maximum achieved 1 kV/mm value). The observed relaxation time versus temperature are summarized in Fig. \ref{fig5}, for each field condition (on or off). Bulk water data from Ref. \cite{Teixeira:85} are shown for comparison. The relaxation times of the hydration layer dynamics, as observed in the H$_2$O-hydrated protein are larger than in the bulk at high temperatures (certainly for $T>250 K$), consistent with the known idea that confinement tends to suppress dynamics. Below 250 K, the relaxation times for the faster component starts  to overlap with that of the bulk liquid, indicating dynamics with comparable time scales .  Fig. \ref{fig5} conveys a key message; that is the applied field of 1 kV/mm does not alter the relaxation times of the adsorbed water in any meaningful way, in contrast to our recent observations in water confined in silica pores \cite{Diallo:12}. The inset figure compares the present observed fast relaxation time $\tau$  (open and solid squares) with those of water adsorbed in tight silica pores \cite{Diallo:12}. We can not definitely rule out the possibility of field-induced enhanced diffusion in proteins because the maximum field achieved in the present study is just 40\% of that reached in the hydrated silica experiment. We note here that the electrical breakdown which sets this maximum field is highly sensitive to factors such as cell geometry and assembly, sample dielectric strength, defects, hydration, surrounding pressure and so on. A small change in any of these factors can have significant impact on the maximum attainable field. We found in our case that no matter how meticulously we prepare our sample assembly, we always find that the protein+water  sample breaks down at a much lower field strength (around 1 kV/mm) than in the silica case, at comparable hydration level.  This is clearly due to a lower dielectric strength of our sample compared to the silica+water system. 

To test the universality of an electric field induced enhanced water diffusion, and to confirm the existence of an onset field value $E_c$,  further work (both MD and experiments) are needed. We anticipate however the intrinsic $E_c$ of confined water to be in the range 2-3 kV/mm, in concordance with our observations in hydrated silica. Field dependence studies of the QENS signal of water adsorbed in other less polarizable proteins (such as Myelin basic protein), would be key in determining the onset field value $E_c$. A key scientific goal is to clarify the mechanism by which this enhanced diffusion takes place at the molecular level, and how it is modified by the substrate interaction. Recent analytical theory and calculations have investigated the dipolar response in various hydrated proteins \cite{Matyushov:12}. In those studies, which included lysozyme, ubiquitin, and cytochrome {\it C} and {\it B}, Matyushov found a remarkable variation of the dielectric constant  between the different proteins. Of particular relevance to our work, it also indicates a strong influence of the coupling of the protein charge surface to the hydration water on the protein overall dipolar response. The protein-water dipolar correlations extend to large distances in all non-neutral proteins and cannot be neglected when evaluating the total dipolar response. In ubiquitin, the only neutral protein of all, the protein self-correlations nearly cancel out the protein-water correlations. We thus anticipate the net effect of field on hydrated ubiquitin to come largely from the hydration layer. This makes  ubiquitin an excellent substrate candidate for future measurements, with the caveats that this protein can be deuterated to yield an observable QENS signal of the hydration water.

\begin{table}% table caption is above the table
{\large
\caption{Temperature dependence of the observed relaxation time $\tau_{1,2}$ with and without field, obtained from the hydration water. The subscripts 1 and 2 refer to the fast and slow diffusion processes associated with the hydration shell, respectively.}
\label{tab:1}       % Give a unique label
\begin{tabular}{cccc}
\hline\noalign{\smallskip}
\tableheadseprule\noalign{\smallskip}
$T$ (K)&$E$ (kV/mm)& $\tau_1$ (ps) & $\tau_2$ (ps)\\
\tableheadseprule\noalign{\smallskip}
220 &0& 38 & 261 \\
%       &1 & - & -\\
\tableheadseprule\noalign{\smallskip}
230 &0 & 24.5 & 177 \\
        & 1&   26.5  &  239\\
\tableheadseprule\noalign{\smallskip}
240 &0&  19.5 &  162 \\
        &1 &   20 & 194\\
\tableheadseprule\noalign{\smallskip}
250 &0&  19.2 & 171 \\
  \tableheadseprule\noalign{\smallskip}
270 &0&  18.7 &131\\
\tableheadseprule\noalign{\smallskip}
   \noalign{\smallskip}\hline
\end{tabular}
}
\end{table}

\section{Conclusions}

We have investigated the impact of  static electric field on the dynamics of lysozyme and  its hydration water. Our aim was to probe the lysozyme response, a highly polar molecule, and to test the universality of the enhanced diffusivity of water under field, observed with silica substrate \cite{Diallo:12}. Our measurements reveal that the nano- to pico-second dynamics of the protein are unaffected by the field, due possibly to the stronger intra-molecular interactions compared to the maximum achieved field strength of 1 kV/mm.  There is also no appreciable quantitative enhancement of the diffusive dynamics of the hydration water, as compared to our observations with water in silica pores in which a field of $~2.5$ kV/mm was achieved.  Within the temperature range investigated here, we see no evidence of a dynamical liquid-liquid transition, as observed by Chen and others \cite{Liu:05,Mamontov:05,Liu:06}. Further measurements, specially at low temperatures, would be necessary to shed light on this very interesting but much debated topic \cite{Liu:05,Liu:06,Doster:10}.

While we argue that the non-observation of any field dependence of the  water dynamics is associated with the difference in interaction potentials of water and the two substrates (protein and silica), or possibly to the existence of a critical threshold field in the range $E_c\sim2-3$ kV/mm,  a conclusive explanation would require complementary MD simulations of the hydration water in lysozyme under comparable field strengths. RecentMD  work  on hydrated proteins at zero field \cite{Hong:11,Qvist:11,Khodadahi:11} could be used as a reference in tackling the field dependence case.

 Meanwhile, we are investigating the possibility of performing additional measurements at higher field values using a highly insulating protein to breakthrough the 1 kV/mm barrier observed here. It also possible for static electric fields  to be screened (Debye screening) due to salt in the protein crystals. Higher purity lysozyme samples in which residual salts in the commercially purchased lysozyme sample are further  removed, could potentially improve the maximum field value for future neutron measurements.

\begin{acknowledgements}
We acknowledge the use of the DAVE software in part of the data analysis \cite{Azuah:09}.  We thank C. Stanley at ORNL for stimulating discussions. It is also pleasure to acknowledge R. Goyette, R. Mills, D. Maierhafer, R. Moody, and M. Loguillo at SNS for valuable technical support. PF acknowledges the GEM fellowship program at UTK. HON and QZ acknowledge the support of the Center for Structural Molecular Biology at ORNL supported by the U.S. DOE, Office of Science, Office of Biological and Environmental Research Project ERKP291. Work at ORNL and SNS is sponsored by the Scientific User Facilities Division, Office of Basic Energy Sciences, U.S. DOE. 
\end{acknowledgements}

\end{document}